\documentstyle[twoside,fleqn,espcrc2,epsf]{article}
\def\centeron#1#2{{\setbox0=\hbox{#1}\setbox1=\hbox{#2}\ifdim
\wd1>\wd0\kern.5\wd1\kern-.5\wd0\fi
\copy0\kern-.5\wd0\kern-.5\wd1\copy1\ifdim\wd0>\wd1
\kern.5\wd0\kern-.5\wd1\fi}}
\def\centerover#1#2{\centeron{#1}{\setbox0=\hbox{#1}\setbox
1=\hbox{#2}\raise\ht0\hbox{\raise\dp1\hbox{\copy1}}}}
\def\centerunder#1#2{\centeron{#1}{\setbox0=\hbox{#1}\setbox
1=\hbox{#2}\lower\dp0\hbox{\lower\ht1\hbox{\copy1}}}}
\def\lsim{\;\centeron{\raise.35ex\hbox{$<$}}{\lower.65ex\hbox
{$\sim$}}\;}
\def\gsim{\;\centeron{\raise.35ex\hbox{$>$}}{\lower.65ex\hbox
{$\sim$}}\;}
\def\nskip#1{\vglue-\baselineskip\vglue#1\vglue-\parskip\noindent}
\title{\nskip{-2.95truecm} \rightline{
\vbox{\normalsize
\halign{&## \hfil\cr
&ANL-HEP-CP-94-91\cr
&November, 1994\cr}
}}\nskip {2.70truecm}%
Decays rates for S- and P-wave bottomonium}
\author{G.~T.~Bodwin, S.~Kim and D.~K.~Sinclair
        \thanks{Talk presented by D.~K.~Sinclair at LATTICE'94, Bielefeld,
                Germany, 27th September -- 1st October, 1994.}
        \thanks{This work was supported by the U.~S. Department of Energy,
                Contract No. W-31-109-ENG-38.}
        \address{HEP Division, Argonne National Laboratory,
                 9700 South Cass Avenue, Argonne, Illinois, 60439, USA}
        }
\begin{document}

\begin{abstract}
We use the Bodwin-Braaten-Lepage factorization scheme to separate the long-
and short-distance factors that contribute to the decay rates of $\Upsilon$,
$\eta_b$ (S-wave) and $\chi_b$,$h_b$ (P-wave). The long distance matrix
elements are calculated on the lattice in the quenched approximation using
a non-relativistic formulation of the $b$ quark dynamics.
\end{abstract}

\maketitle

In heavy quarkonium decays that involve quark-antiquark ($Q\bar{Q}$)
annihilation, this annihilation occurs at short distances ($\sim 1/M_q$).
Bodwin, Braaten and Lepage \cite{BBL} have shown that this enables one to
factor such decay rates into a sum of products of a short-distance
parton-level decay rate with a long-distance matrix element between quarkonium
states. The short distance pieces are calculated perturbatively, while the long
distance parts are accessible to lattice calculations. To lowest non-trivial
order in $v^2$, the square of the quark velocity, ($v^2 \sim .1$ for
bottomonium)
\begin{eqnarray}
\Gamma(^{2s+1}S_J\rightarrow X)\!\!\!\!\!\!\!\!\!\!\!
&&=G_1(S)\hat{\Gamma}_1(Q\bar{Q}(^{2s+1}S_J)\rightarrow X) \nonumber \\
\Gamma(^{2s+1}P_J\rightarrow X)\!\!\!\!\!\!\!\!\!\!\!
&&=H_1(P)\hat{\Gamma}_1(Q\bar{Q}(^{2s+1}P_J)\rightarrow X) \nonumber \\
&&+\; H_8(P)\hat{\Gamma}_8(Q\bar{Q}(^{2s+1}S_J)\rightarrow X),\nonumber \\
&&
\end{eqnarray}
where the $X$'s represent states of light partons. The $\hat{\Gamma}$'s
are the short-distance ($p \sim M_Q$) parton-level decay rates. $G_1$, $H_1$
and $H_8$ are the long-distance ($p \sim M_Qv$, $E \sim M_Qv^2$) matrix
elements that we calculate on the lattice.

In our lattice calculations we have used 149 independent equilibrated quenched
gauge configurations on a $16^3 \times 32$ lattice with $\beta=6.0$. Heavy-
quark, and hence quarkonium, propagators were calculated using the
non-relativistic formulation of Lepage and collaborators \cite{lepage}. We used
the lattice version of the quark action that is based on the euclidean
lagrangian
\begin{equation}
{\cal L}_Q = \psi^{\dag}(D_t-{{\bf D}^2 \over 2M_Q})\psi+
             \chi^{\dag}(D_t+{{\bf D}^2 \over 2M_Q})\chi,
\end{equation}
which is valid to the lowest non-trivial order in $v^2$. We calculate the
quark Green's function that obeys the evolution equation \cite{lepage}
\begin{eqnarray}
G({\bf x},t+1)\!\!\!\!\!\!\!\!\!\!\!
&&=(1-H_0/2n)^n U_{{\bf x},t}^{\dag} (1-H_0/2n)^n  \nonumber \\
&&\times G({\bf x},t)+\delta_{\bf x,0}\delta_{t+1,0} \; ,
\end{eqnarray}
with $G({\bf x},t)=0$ for $t < 0$, and $H_0=-\Delta^{(2)}/2M_0-h_0$ .
Here $\Delta^{(2)}$ is the gauge-covariant discrete laplacian, and $M_0$ the
bare quark mass. $h_0=3(1-u_0)/M_0$, where
$u_0=\langle 0 |\frac{1}{3}{\rm Tr} U_{plaq} | 0 \rangle^{\frac{1}{4}}$.

The matrix elements we calculate are defined as
\begin{equation}
G_1=\langle ^1S|\psi^{\dag}\chi\chi^{\dag}\psi|^1S\rangle/M_Q^2
\end{equation}
\begin{equation}
H_1=\langle ^1P|\psi^{\dag}(i/2)\stackrel{\leftrightarrow}{\bf D}\chi.
\chi^{\dag}(i/2)\stackrel{\leftrightarrow}{\bf D}\psi|^1P\rangle/M_Q^4
\end{equation}
\begin{equation}
H_8=\langle ^1P|\psi^{\dag}T^a\chi\chi^{\dag}T^a\psi|^1P\rangle/M_Q^2 \; ,
\end{equation}
On the lattice, we calculate the related quantities $G^\ast_1$, $H^\ast_1$,
$H^\ast_8$, defined graphically below
\centerline{
\begin{picture}(200,125)
\put(50,95){\oval(80,40)}
\put(140,95){\oval(100,40)}
\put(0,65){\line(1,0){200}}
\put(47.5,35){\oval(75,40)}
\put(142.5,35){\oval(95,40)}
\put(10,95){\circle*{10}}
\put(190,95){\circle*{10}}
\put(10,35){\circle*{10}}
\put(190,35){\circle*{10}}
\put(90,95){\circle*{5}}
\put(85,35){\circle*{5}}
\put(95,35){\circle*{5}}
\put(45,95){\vector(-1,0){30}}
\put(55,95){\vector(1,0){32.5}}
\put(135,95){\vector(-1,0){42.5}}
\put(145,95){\vector(1,0){40}}
\put(47,92){T}
\put(137,92){T'}
\put(195,10){,}
\end{picture}
}
where the larger dots represent the ``sources'', the small dot in the numerator
is the appropriate 4-fermi operator, and the small dots in the denominator
represent point ``sinks''. For our calculations we generate the retarded
(Eqn.~(3)) and advanced quark propagators from noisy point and noisy extended
sources on each of the 32 time-slices. (This differs from our preliminary
calculations, in which the 4-fermi operator was used as a source.) Then as
$T,T'\rightarrow\infty$
\begin{equation}
G_1^\ast(T,T') \rightarrow G_1{2\pi M_Q^2 \over 3|R_{1S}(0)|^2}=1+{\cal O}(v^4)
\end{equation}
\begin{equation}
H_1^\ast(T,T') \rightarrow H_1{2\pi M_Q^4 \over 9|R'_{1P}(0)|^2}=1+{\cal
O}(v^4)
\end{equation}
\begin{equation}
H_8^\ast(T,T') \rightarrow  {H_8 \over M_Q^2 H_1}+{\cal O}(v^4) \; ,
\end{equation}
where $R_{1S}$ is the radial wave function of the 1S state and $R'_{1P}$ is the
derivative of the radial wave function of the 1P state.

\begin{figure}[htb]
\vspace{-1in}
\epsfxsize=4in
\epsffile{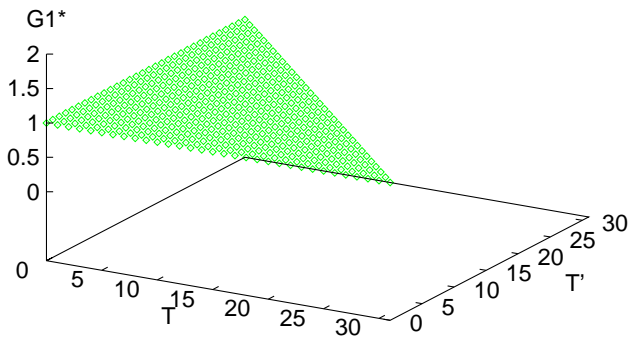}
\caption{$G_1^\ast$ as a function of $T$ and $T'$.}
\label{fig:g1}
\end{figure}

\begin{figure}[htb]
\vspace{-1in}
\epsfxsize=4in
\epsffile{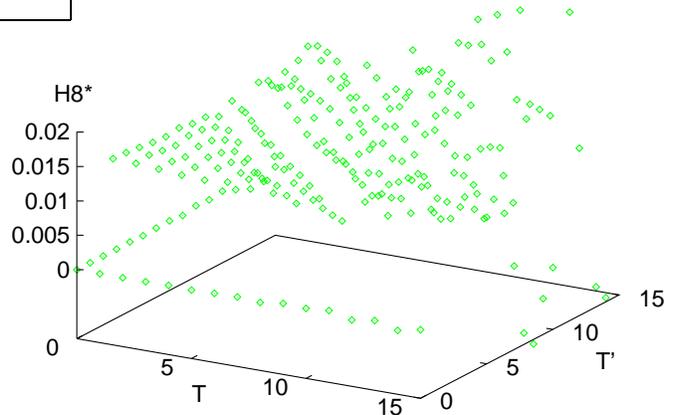}
\caption{$H_8^\ast$ as a function of $T$ and $T'$.}
\label{fig:h8}
\end{figure}

For bottomonium, we use input parameters determined by the NRQCD collaboration
\cite{NRQCD}, which in our convention are: the bare b-quark mass, $M_{0b}=1.5$,
the inverse lattice spacing, $a^{-1}=2.4$GeV, and the physical b-quark mass,
$M_b=2.06$. In Fig.~\ref{fig:g1} we show $G^\ast_1$ as a function of $T$,$T'$.
It is clearly very close to the vacuum saturation value of 1. In fact
$G^\ast_1 - 1 \approx 1.3 \times 10^{-3}$. $H^\ast_1$ displays similar
behaviour, but is more noisy. $H^\ast_8$ is plotted in Fig.~\ref{fig:h8}. We
notice that it displays a fairly obvious plateau at small $T$, $T'$, which
degenerates into noise for larger values of $T$, $T'$. No improvement in
$H^\ast_8$ is obtained by using the extended source. Fitting the plateau, we
obtain
$$
          {H_8/H_1} \approx 0.06 \; .
$$
This is somewhat smaller than the value obtained from a simple perturbative
estimate \cite{BBL}. However, this estimate comes from assuming that $H_8$
becomes negligibly small when the momentum cutoff is $\Lambda_{QCD}$. If one
assumes, instead, that $H_8$ becomes negligible at a cutoff closer to the
bottomonium binding energy, then the perturbative estimate is closer to the
lattice measurement. Of course, the lattice-regulated $G_1$, $H_1$ and $H_8$
differ from their continuum counterparts at ${\cal O}(\alpha_s)$, but since our
methods are equivalent to using mean-field improved actions, these
renormalizations are expected to be small.

We have also considered the S-wave decays through next-to-leading order in
$v^2$. To this order, $G_1$ is no longer the same for $\Upsilon$ and $\eta_b$.
However, we would need an improved action in order to calculate these
corrections. In addition, there is a second term in Eqn.~(1),
\begin{equation}
     F_1(S)\hat{\Gamma}'_1(Q\bar{Q}(^{2s+1}S_J \rightarrow X) \; ,
\end{equation}
where $\hat{\Gamma}'_1$ is another perturbative parton-level decay rate and
$F_1$ can be calculated on the lattice using the Lagrangian of Eqn.~(2). In
the vacuum saturation approximation,
\begin{equation}
     F_1(S) = \langle 0 |\psi^{\dag}\chi|0 \rangle  \langle 0 |\psi^{\dag}
    (\frac{-i}{2}\stackrel{\leftrightarrow}{\bf D})^2\chi| 0 \rangle/M_Q^4 \; .
\end{equation}
On the lattice we measure $F^\ast_1$, defined as
\begin{equation}
     F^\ast_1 = {M_Q^2 F_1 \over G_1} \; .
\end{equation}
We find that
\begin{eqnarray}
      F^\ast_1 & = & 1.3134(9) \hbox{ --- non-covariant} \\
      F^\ast_1 & = & 0.8519(6) \hbox{ --- covariant} \; ,
\end{eqnarray}
where non-covariant and covariant refer to whether we use ordinary derivatives
(in coulomb gauge) or gauge-covariant derivatives in Eqn. (11). As with $G_1$,
$H_1$ and $H_8$, $F^\ast_1$ requires renormalization. $F_1$ mixes with $G_1$.
Since $F_1/G_1 \sim v^2$, this mixing can be significant. We have calculated
these mixings to 1-loop order. Preliminary estimates of the $F^\ast_1$'s which
take these mixings into account are
\begin{eqnarray}
      F^\ast_1(renormalized)\!\!\! & = &\!\!\! 0.76 \hbox{ --- non-covariant}
\\
      F^\ast_1(renormalized)\!\!\! & = &\!\!\! 0.62 \hbox{ --- covariant} \; .
\end{eqnarray}

Finally, in table~\ref{tab:bottom} we present some mass and wavefunction
calculations which were incidental to our calculations of matrix elements.
Clearly our numbers are inferior to those obtained by the NRQCD collaboration
\cite{NRQCD}, since we work only to lowest non-trivial order in $v^2$. However,
they serve as a consistency check of our calculations.

We are now in the process of repeating these calculations for the charmonium
system at $\beta=5.7$ ($\beta=6.0$ has too small a lattice spacing for NRQCD at
the charmed-quark mass. Our earlier attempts \cite{BKS} used charmed-quark
masses which were too large.) The charmonium system affords the opportunity to
confront our calculations with experiment, since there is already sufficient
experimental data to allow extraction of $H_8$. In the future, we hope to
extend these calculations to next order in $v^2$ and $a^2$, and then to include
the effects of light dynamical quarks. We are also calculating the complete
renormalization matrix through ${\cal O}(\alpha_s)$ for the four operators
discussed in this paper.
\begin{table*}[htb]
\setlength{\tabcolsep}{0.25pc}
\caption{Properties of S- and P-wave bottomonium from our simulations.
         The lattice quantities include mean field renormalizations.
         The mass of the 1S state is obtained by using $M=2(Z_M M_b - E_0)+E_n$
         with $Z_M$ and $E_0$ set at their mean field values.}
\label{tab:bottom}
\begin{tabular}{|l|c|c|}
\hline
                 & LATTICE               & EXPERIMENT        \\
\hline
$M_{1S}$         & 9.2766(9) GeV         & $M_\Upsilon$ = 9.46037(21) GeV \\
$M_{1P}-M_{1S}$  & 0.434(9) GeV          & $M_{\chi_b}-M_\Upsilon$
                                             = 0.4398(7) GeV \\
$|R_{1S}(0)|^2$  & 4.33(2) GeV$^3$       & 7.2(2) GeV$^3$    \\
$|R'_{1P}(0)|^2$ & 0.75(7) GeV$^5$       &     ---           \\
\hline
\end{tabular}
\end{table*}

We wish to thank John Sloan, Christine Davies and G.~Peter Lepage for
informative discussions and insight. We thank J.~B.~Kogut, M.-P.~Lombardo and
D.~K.~Sinclair for allowing us to use their gauge configurations. These
calculations were performed on the CRAY C-90 at NERSC, whose resources were
made available to us through the Energy Research Division of the U.S.
Department
of Energy.

\end{document}